\documentclass[twocolumn]{aastex631}

\usepackage{threeparttable}

\usepackage[titletoc]{appendix}
\usepackage[T1]{fontenc}
\usepackage{CJK}

\usepackage{graphicx}	% Including figure files
\usepackage{amsmath}	% Advanced maths commands
\usepackage{float}
\usepackage{subfigure}
\usepackage{color}
\shorttitle{Stellar Parameters for Gaia Cool Dwarfs}
\shortauthors{Cai-Xia Qu et al.}

\begin{document}

\title{Stellar Atmospheric Parameters for Cool Dwarfs in Gaia DR3}

\correspondingauthor{A-Li Luo}
\email{lal@nao.cas.cn}

\author[0000-0002-5460-2205]{Cai-Xia Qu}
\affiliation{CAS Key Laboratory of Optical Astronomy, National Astronomical Observatories, Chinese Academy of Sciences, Beijing 100101, China}
\affiliation{School of Astronomy and Space Science, University of Chinese Academy of Sciences, Beijing 100049, China}

\author[0000-0001-7865-2648]{A-Li Luo}
\affiliation{CAS Key Laboratory of Optical Astronomy, National Astronomical Observatories, Chinese Academy of Sciences, Beijing 100101, China}
\affiliation{School of Astronomy and Space Science, University of Chinese Academy of Sciences, Beijing 100049, China}

\author[0000-0001-6767-2395]{Rui Wang}
\affiliation{CAS Key Laboratory of Optical Astronomy, National Astronomical Observatories, Chinese Academy of Sciences, Beijing 100101, China}
\affiliation{School of Astronomy and Space Science, University of Chinese Academy of Sciences, Beijing 100049, China}

\author[0000-0003-0433-3665]{Hugh R. A. Jones}
\affiliation{School of Physics, Astronomy and Mathematics, University of Hertfordshire, College Lane, Hatfield AL10 9AB, UK}

\author[0000-0003-0716-1029]{Bing Du}
\affiliation{CAS Key Laboratory of Optical Astronomy, National Astronomical Observatories, Chinese Academy of Sciences, Beijing 100101, China}
\affiliation{School of Astronomy and Space Science, University of Chinese Academy of Sciences, Beijing 100049, China}

\author[0000-0001-5738-9625]{Xiang-Lei Chen}
\affiliation{CAS Key Laboratory of Optical Astronomy, National Astronomical Observatories, Chinese Academy of Sciences, Beijing 100101, China}
\affiliation{School of Astronomy and Space Science, University of Chinese Academy of Sciences, Beijing 100049, China}

\author[0000-0001-7671-4745]{You-Fen Wang}
\affiliation{CAS Key Laboratory of Optical Astronomy, National Astronomical Observatories, Chinese Academy of Sciences, Beijing 100101, China}
\affiliation{School of Astronomy and Space Science, University of Chinese Academy of Sciences, Beijing 100049, China}

%% Note that the \and command from previous versions of AASTeX is now
%% depreciated in this version as it is no longer necessary. AASTeX 
%% automatically takes care of all commas and "and"s between authors names.

%% AASTeX 6.31 has the new \collaboration and \nocollaboration commands to
%% provide the collaboration status of a group of authors. These commands 
%% can be used either before or after the list of corresponding authors. The
%% argument for \collaboration is the collaboration identifier. Authors are
%% encouraged to surround collaboration identifiers with ()s. The 
%% \nocollaboration command takes no argument and exists to indicate that
%% the nearby authors are not part of surrounding collaborations.

%% Mark off the abstract in the ``abstract'' environment. 
\begin{abstract}
We provide a catalogue of atmospheric parameters for 1,806,921 cool dwarfs from Gaia DR3 which lie within the range covered by LAMOST cool dwarf spectroscopic parameters: 3200 K $<T_{eff}<$ 4300 K, -0.8 $< [M/H] <$ 0.2 dex, and 4.5 $<log \emph{g} <$ 5.5 dex. Our values are derived based on Machine Learning models trained with multi-band photometry corrected for dust. The photometric data comprises of optical from SDSS r, i, z bands, near-infrared from 2MASS J, H, K and mid-infrared from ALLWISE W1, W2. We used both random forest and LightGBM machine learning models and found similar results from both with an error dispersion of 68 K, 0.22 dex, and 0.05 dex for $T_{eff}$, [M/H], and log $\emph{g}$, respectively. Assessment of the relative feature importance of different photometric colors indicated W1 -- W2 as most sensitive to both $T_{eff}$ and log \emph{g}, with J -- H most sensitive to [M/H]. We find that our values show a good agreement with APOGEE, but are significantly different to those provided as part of Gaia DR3. 
\end{abstract}


\section{INTRODUCTION} \label{sec:INTRODUCTION}
Cool dwarfs with low mass and low luminosity constitute more than $70\%$ of objects in the Galaxy. Determination of their stellar atmosphere parameters (APs) is vital for exploring the stellar formation, composition, and evolution history of the Galaxy (e.g., \cite{2007AJ....134.2418B}). However, the estimation of APs of cool dwarfs is difficult because of their complex atmosphere along with convective mixing. With ongoing improvements in atmosphere modeling for low-mass stars and advancements in the number of observations made and variety of  instruments used, the measurement of the APs of cool stars has been carried out increasingly precisely.

Observations such as \cite{1996MNRAS.280...77J} used the PHOENIX synthetic spectra to infer the APs of a few M dwarfs by comparing them with observed spectra. \cite{2008MNRAS.389..585C} obtained the effective temperatures ($T_{eff}$) and bolometric luminosity of the M dwarfs based on the empirical relationship between the flux ratio in different bands and both the $T_{eff}$ and the metallicity. \cite{2021RAA....21..202D} estimated the atmospheric parameters of M-type stars using an updated pipeline LASPM from low-resolution spectra, while \cite{2022ApJS..260...45D} applied the ULySS package with MILES interpolator model spectra to estimate the APs of M dwarfs. Furthermore, \cite{2021ApJS..253...45L} published a stellar AP catalog of LAMOST M dwarfs using the SLAM machine learning algorithm \citep{2020ApJS..246....9Z}. 

The increasing availability of near-infrared spectra has also been utilized to determine precise APs of M dwarfs (e.g.,\cite{2010ApJ...720L.113R, 2022MNRAS.511.1893C, 2022arXiv220205858H}). \cite{2012ApJ...748...93R} calibrated the H$_2$O - K2 index of K-band spectra of M dwarfs and estimated $T_{eff}$ and [M/H] of these objects based on NaI, CaI, and H$_2$O - K2 indexes.

\begin{figure*}[htb!]
    \centering
    \includegraphics[width=0.97\textwidth]{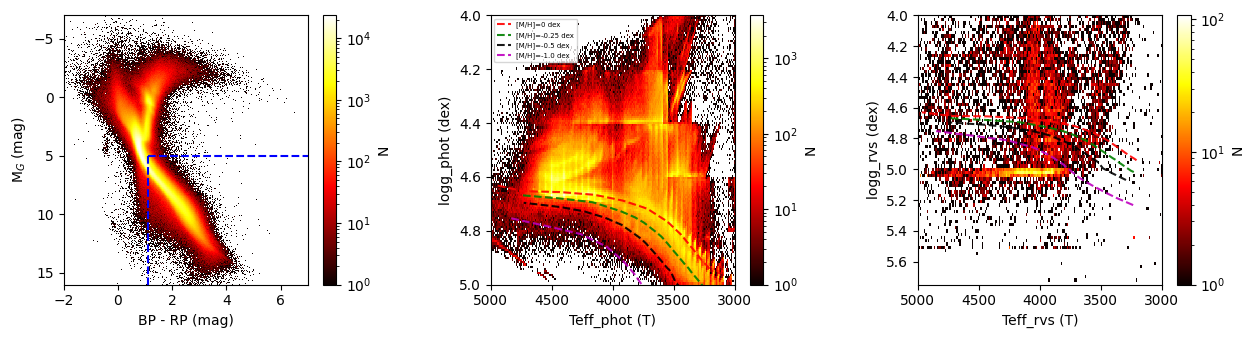}
    \caption{Distributions of Gaia objects and their APs. The left panel represents the HRD of Gaia DR3 objects. The blue dashed lines provide preliminary bounds for our sub-sample of 'cool dwarfs' based on the position of vertical and horizontal blue lines at (1.3, 5). The middle panel plots the Kiel diagram (T$_{eff}$ vs log $\emph{g}$) based on parameters assigned using BP/RP spectra of the objects selected by the blue dashed lines in the left panel from \cite{2023A&A...674A..27A}, as well as PARSEC isochrones with an age of 6 Gyr and [M/H] of 0, -0.25, -0.5 and -1.0 dex. The right panel depicts the same Kiel diagram for objects with parameters based on RVS spectra from \cite{2023A&A...674A..29R} along with the same isochrones used in the middle panel.}
    \label{fig:gaia_kiel}
 \end{figure*}
 
Gaia is a satellite launched by the European Space Agency, with the goal of providing precise 3D maps and space motions of approximately one billion stars in our Galaxy \citep{2016A&A...595A...1G,2016AA...595A...2G, 2018A&A...616A...5C}. In 2022, Gaia Data Release 3 (DR3) published complete data products \citep{2022arXiv220800211G}, which include photometry in G, G$_{BP}$, G$_{RP}$, objects with various types, mean low-resolution (BP/RP) spectra and high-resolution Radial Velocity Spectrometer (RVS) spectra. With the BP/RP spectra, APs of 470,759,263 sources within $G < 19$ mag were measured \citep{2023A&A...674A..27A}, while with RVS spectra,  APs of 5,591,594 objects were measured, most of which are AFGK stars \citep{2023A&A...674A..29R}. Figure \ref{fig:gaia_kiel} shows diagrams with the distribution of Gaia DR3 cool dwarfs. The left panel is the Hertzsprung–Russell diagram (HRD) of all Gaia DR3 color-coded by number density. The objects below the blue dashed lines are considered to be cool dwarfs, whose Kiel diagrams (log $\emph{g}$ vs. T$_{eff}$) obtained from BP/RP and RVS spectra are shown in the middle and right panels, respectively. 

The APs distributions shown in the middle and right plots of Figure \ref{fig:gaia_kiel} are not consistent with each other. Nor are they consistent with the expectation for these dwarfs to be objects to largely be on the main sequence and so approximately overlapping with standard isochrones. There are known problems with the external calibration of BP/RP spectra for cool objects which may impact the middle plot, e.g., \cite{2023A&A...669A.139S}. Given that cool dwarfs are relatively faint and the narrow band of RVS spectra we might anticipate the issues with the derivation of parameters in the right-hand plot which indicate the distinct selection of log $g$=5.0 as well as vertical features in temperature selection which appear un-physical. We note the AP estimation using Gaia BP/RP spectra by \cite{2023MNRAS.tmp.1869Z} who develop a data-driven model to estimate the APs of Gaia 220 million objects though limited to T$_{eff}$ > 4000 K.

High-resolution spectroscopy can provide precise stellar APs but the number of objects that can be observed is limited. The high-resolution survey SDSS/APOGEE DR16\footnote{https://www.sdss.org/dr16/irspec/parameters/} published 22,991 cool dwarfs benefiting from infrared spectra that suffer less from line blending in the inferences of APs than optical spectra. However, an ongoing problem is the difference between the synthetic and observed spectra. \cite{2016A&A...587A..19P} used new synthetic spectra based on the PHOENIX grid to determine APs from high-resolution spectroscopic observations. However, the [Fe/H] reported from earlier estimations in some cases disagree with the new work by more than $3\sigma$. The degeneracy of stellar APs is a significant issue.
\cite{2018A&A...620A.180R} determined $T_{eff}$, [Fe/H], and log $\emph{g}$ for 292 high-resolution spectra using the BT-Settl model although the $\chi ^{2}$ map in figure 3 of \citep{2018A&A...620A.180R} shows degeneracy between different parameter combinations.

On the other hand low-resolution surveys can collect hundreds of thousands of cool dwarf spectra which should be representative of the larger sample of M dwarfs. LAMOST DR9\footnote{http://www.lamost.org/dr9/v1.0/doc/lr-data-production-description} has released a stellar AP catalog of M stars, comprising more than 0.6 million M dwarfs and benchmarked against other sources. Here we use these as the ground truth for machine learning to derive APs of Gaia DR3 CDs with multi-band photometry. 

Multiband photometry provides a robust sampling of the spectra energy distribution to compensate for the lack of local information in spectra. Additionally, selection effects can be avoided when using multiband photometry. Compared to spectroscopy, multiband photometry can ensure the accuracy of AP estimation, although its precision maybe relatively low. Furthermore, photometry can detect fainter objects and provide a more consistent data distribution. In recent years, several ground- and space-based surveys have released a vast amount of photometry in the Milky Way, such as 2MASS \citep{2006AJ....131.1163S}, WISE \citep{2010AJ....140.1868W}, Pan-StaRRS \cite{2002SPIE.4836..154K}, SkyMapper \cite{2011PASP..123..789B}, SAGE \citep{2018RAA....18..147Z}, SAGA \citep{2014ApJ...787..110C}, J-PLASS \citep{2021arXiv211207304Y}, Gaia \citep{2016AA...595A...2G} and SDSS \citep{2000AJ....120.1579Y}). 

%Many authors have used these large databases of photometry to provide APs. For example, \cite{1996AA...313..873A} fitted a polynomial function of $T_{eff}$ versus colors ((B - V), (R - I), (V - R), (V - I), (V - K), (J - H), (J - K), and ubvy-$\beta$), in which $T_{eff}$ is from the InfraRed Flux Method with the new grid of atmosphere models developed by \cite{1993sssp.book.....K}. \cite{2021MNRAS.508.5148C} presented a photometric ( J - H color) and high-resolution spectroscopic calibration of $T_{eff}$, [Fe/H], and log $\emph{g}$ for M dwarfs based on the principal component analysis (PCA). \cite{2016MNRAS.460.2611S} displayed an empirical relation for $T_{eff}$ as a function of r - z, combined with the W1 - W2 color, achieving a precision of 100 K in $T_{eff}$ from colors alone.

Taking advantage of multiband photometry and machine learning methods, we use optical and infrared photometry along with APs dervied from LAMOST spectra to infer the stellar APs for cool dwarfs. Two different machine learning (ML) algorithms were used to train models for APs to ensure consistency and performance when testing and deriving parameters. The paper is organized as follows. Section \ref{sec:data} presents a detailed description of the cool dwarf sample, photometry surveys, and the data processes for the training sample. Section \ref{sec:model} describes the machine learning algorithms, model construction of APs, and feature importance analysis. Section \ref{sec:catalog} presents an AP catalog of cool dwarfs in Gaia DR3. Finally, the main works are summarized in Section \ref{sec:summary}.

\section{Data} \label{sec:data}
In this section, we select cool dwarfs from Gaia DR3 to estimate their APs. Multiband photometry was used as input for our ML algorithms. The training sample was selected from LAMOST M dwarfs that had precise APs from two previous works.

\subsection{Cool Dwarfs Sample Selection}

\begin{figure}[h]
    \centering
    \includegraphics[width=0.45\textwidth]{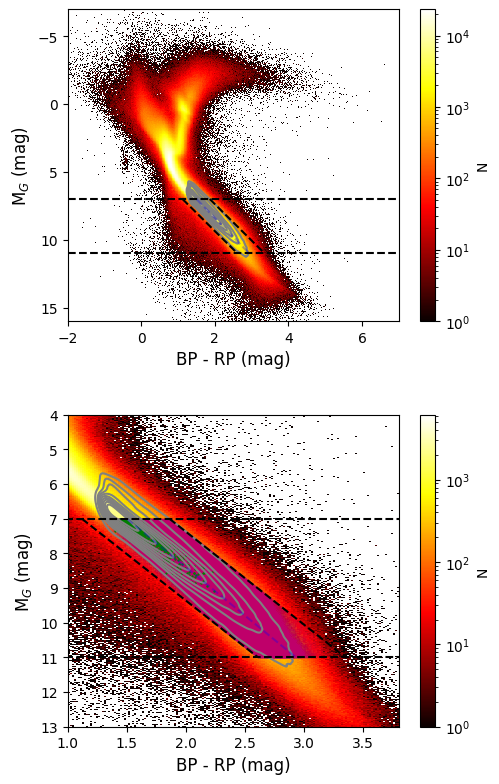}
    \caption{Color-magnitude diagram of Gaia stars and selected stars. Top: the background displays Gaia DR3 objects, while the training sample is plotted as grey contours with logarithmic scaling and fitted with a blue dashed line. The oblique black dashed lines are created by shifting the blue line up and down 1.0 mag. The two horizontal black lines represent the absolute G magnitude of 7 and 11 mag, respectively. Bottom: a zoom-in panel of the top panel with M$_{G}$ between 4 and 13 mag with the addition of Gaia sources with APs. The green points are the sources with APs from RVS spectra, while the magenta points are the objects with APs from BP/RP spectra that lie within the black dashed selection box.}
    \label{fig:gaia_hrd}
\end{figure}

We select an appropriate sample of cool dwarfs from Gaia DR3 based on the dashed black lines in the color-magnitude diagram presented in Figure \ref{fig:gaia_hrd}. Our rationale for this is based on finding simple but appropriate cuts indicated by the grey contours in the training sample. To construct these dashed black lines we plot the LAMOST training sample presented in Section \ref{sec:traing sample} as a grey contours. First, we performed a linear fit on the training sample, indicated by the blue dashed line in Figure \ref{fig:gaia_hrd}. Then, this blue dashed line then was shifted up and down by a magnitude to create two parallel black dashed lines enclosing the training sample. The magnitude limits for this training sample are set between M$_{G}$=7-11.  These choices are based on noticing the change of contour shape for M$_{G}$ $< $7 mag and that the last training set contour extends only slightly beyond M$_{G}$=11. As shown in both panels of Figure \ref{fig:gaia_hrd}, the box composed of black dashed lines is the selected region. We note that approximately 0.1 million sources with BP - RP > 2.9 mag lie beyond a significant density of training sample sources. For these objects, their APs will have an additional column labeled 'flag' to caution others when utilizing these APs.

The bottom panel of Figure \ref{fig:gaia_hrd}, shows a zoom-in for M$_{G}$ between 4 and 13 mag where the green and magenta objects are the Gaia sources with APs from RVS (587 sources) and BP/RP spectra (1,534,957 sources), respectively. Additionally, we used other criteria to exclude unreliable photometry and exclude unresolved binary stars. These criteria are as follows:

\begin{enumerate}
    \item Gmag/error$\_$Gmag > 20, BPmag/error$\_$BPmag > 20, and RPmag/error$\_$RPmag > 20;
    \item Photometry error in each band less than 0.08 mag;
    \item RUWE < 1.4;
    \item parallax/error$\_$parallax > 5.
\end{enumerate}

The basis for criteria 1 is to enable accurate photometry and astronomy \citep{2021ApJS..253...45L}, as well as criteria 2 and 4 \citep{2022MNRAS.510..433L} and 4. Criteria 3 is for the avoidance of binary stars which are likely to provide blended photometry \citep{2023A&A...674A..39G}, which also will be applied to the training sample.

\begin{figure}[h]
    \centering
    \includegraphics[width=0.45\textwidth]{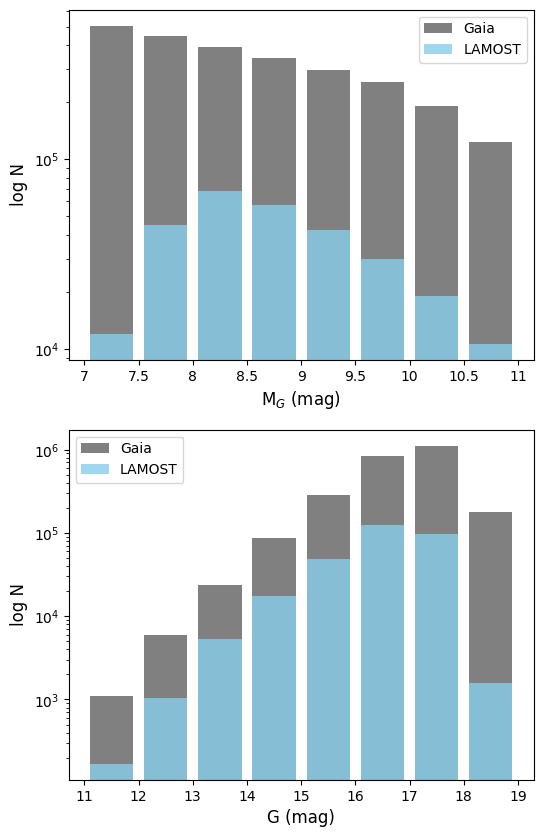}
    \caption{Source number comparison between Gaia cool dwarfs (in grey) and those with spectroscopic APs of LAMOST (in blue) for M$_G$ (top) and G (bottom) photometry.}
    \label{fig:gaia_apo_lamo}
\end{figure}

We thus find 1,806,921 cool dwarfs from Gaia DR3 which overlap with the parameter space of our LAMOST training sample. Compared to spectroscopic data, only a small fraction of these stars have APs already from LAMOST. A comparison of the overlap between the Gaia cool dwarfs and LAMOST samples is shown in Figure \ref{fig:gaia_apo_lamo}. 
%The top and bottom panels are the number with log scale in absolute (M$_G$) and apparent (G) magnitude in G band, respectively. It can be seen that a large numbers of Gaia cool dwarfs do not have APs at fainter magnitudes absolute and apparent magnitudes.

\subsection{Photometric Band Selection}
To connect the APs of the LAMOST-Gaia matching cool dwarfs to those without LAMOST APs we selected the multiband photometry from the optical to the infrared band as input data. Sloan Digital Sky Survey (SDSS) is a 2.5-meter telescope located at the Astronomical Observatory in Apache, New Mexico \citep{2000AJ....120.1579Y}. There are five narrow bands u, g, r, i, and z (centered on 3551\AA, 4686\AA, 6166\AA, 7480\AA, and 8932\AA) ranging from the optical to near-infrared wavelength range. However, many cooler dwarfs are too faint to appear in the u and g band and so only the r, i and z bands are considered. The Two Micron All-Sky Survey (2MASS) is a near-infrared digital imaging survey of the entire sky conducted by the University of Massachusetts and IPAC at 1.25, 1.65, and 2.17 microns \citep{1969tmss.book.....N}. 2MASS can uniformly scan the entire sky in three near-infrared bands, J (1.25 microns), H (1.65 microns), and K (2.17 microns). The Wide-Field Infrared Survey Explorer (WISE) is an all-sky survey from 3 to 25 micrometers that aims to provide vast storage of knowledge about the Solar System, the Milky Way, and the Universe \citep{2010AJ....140.1868W}. It was launched on 14 December 2009 and updated the last data on 21 March 2013. The WISE source catalog contains the attributes of 563,921,584 resolved and point-like objects detected in the Atlas intensity images, with four bands W1, W2, W3, and W4 (3.4, 4.6, 12, and 22 $\mu$ m). To ensure the quality of photometry, we set the error of each band to less than 0.08 mag. We also exclude some possible extended sources by flag 2MASS gal$\_$contam and mp$\_$glag = 0, and flag ALLWISE ext$\_$flag = 0 \citep{2022MNRAS.510..433L}. We obtained a final Gaia cool dwarf sample of 1,806,921 objects by matching these multiband photometry in TOPCAT with a radius of 3''. We do not include G, BP, and RP magnitudes as part of the training set in order to provide a solution which is independent. In particular, we avoid the use of BP and RP since potentially they are part of the discrepancy presented in the middle pot of Figure \ref{fig:gaia_kiel}.

\begin{figure}
	\centering
	\includegraphics[scale=0.5]{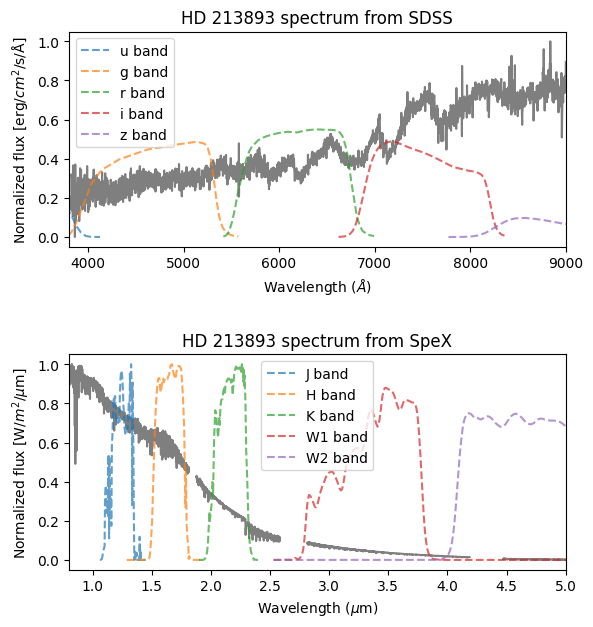}
	\caption{Example spectra of HD 213893 shown from 0.4 to 5 microns. The top panel shows an optical spectrum observed by SDSS, with RSR lines in u, g, r, i, and z bands plotted. The bottom panel shows the infrared spectrum from SpeX, which covers J, H, K, W1, and W2 bands.}
	\label{fig:spec_sample}
\end{figure}

The data features used for this work include photometry in eight bands (r, i, z, J, H, K, W1, and W2). As an example, the spectra of HD 213893 that is classified as M0 type and plotted in Figure \ref{fig:spec_sample} from optical to infrared.  
The top panel shows the optical spectra from SDSS, with relative system response (RSR) curves of five bands provided by the SDSS\footnote{http://classic.sdss.org/dr7/instruments/imager/index.html}. Prominent spectral features appear in the r, i, and z bands such as Na D line and molecular bands (CaH, TiO, VO \citep{2003AJ....125.1598L}).
The bottom panel presents the spectrum with a wavelength of 0.8 to 5.5 $\mu$m and a resolution of R $\approx$ 2000, which is from the 3.0 m NASA Infrared Telescope Facility (IRTF) \citep{2003PASP..115..362R}. The corresponding RSR lines are presented according to the introduction of 2MASS\footnote{https://old.ipac.caltech.edu/2mass/releases/allsky/doc/} and WISE\footnote{https://wise2.ipac.caltech.edu/docs/release/allsky/expsup/}. It can be seen that the flux density is high in the infrared range. Considering the accuracy in W1 and W2 bands of WISE \citep{2013MNRAS.430.2188Y}, we adopted W1 and W2 photometry, although the spectral features are weaker in the far-infrared band. Therefore, photometry in the r, i, z, J, H, K, W1, and W2 bands was employed to the represent the features of our data.

\subsection{Extinction Correction}
The 3D dust map method \citep{Green2018}, consists of two-dimensional maps \citep{2014AA...571A..11P} and three-dimensional maps \citep{2018MNRAS.478..651G, Green_2015}  applied to correct the extinction. In addition to sky position of each point, reliable distance is also needed in 3D dust map method. Here, we used the parallax from Gaia to compute the distance, which is considered to be useful to a distance of 4 kpc and containing all our objects. The extinction coefficients in different bands \citep{2022MNRAS.510..433L, 2018MNRAS.475.5023C, 2018MNRAS.479L.102C, 2014ApJ...787..110C, 2013MNRAS.430.2188Y} were used to transform to the corresponding bands. Furthermore, the absolute magnitudes in multiple bands were also computed.

\subsection{Training Sample} \label{sec:traing sample}
To obtain AP labels of cool dwarfs, we used the recent works of AP estimation of LAMOST M dwarfs, we cross-matched the two catalogs of \cite{2021ApJS..253...45L} (hereafter L21) and \cite{2022ApJS..260...45D} (hereafter D22). In the L21 catalog, the $T_{eff}$ and [M/H] are trained by APOGEE APs. Meanwhile the $T_{eff}$, [M/H], and log \emph{g} are all shown in the D22 catalog, trained by the ULySS package. 

\begin{figure*}[t]
    \centering
	\includegraphics[width=\textwidth]{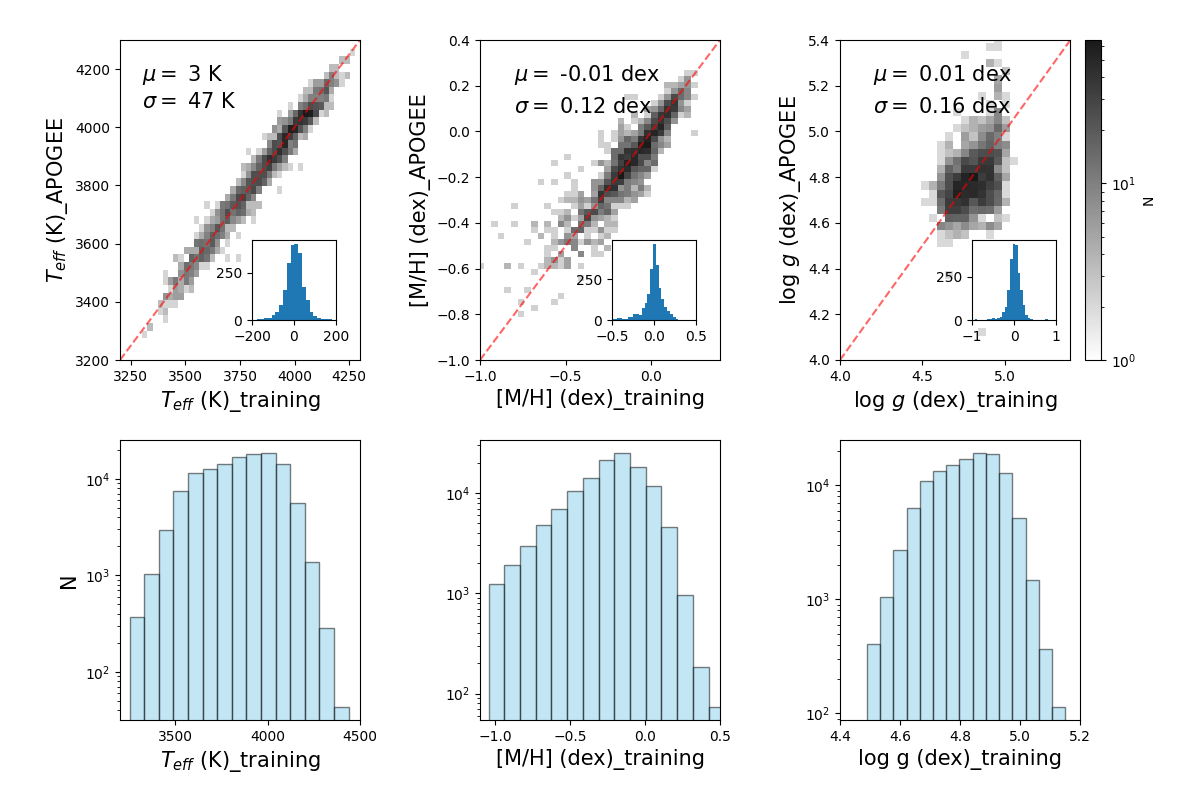}
	\caption{The upper row shows density plots comparing APs of LAMOST training sample with that of APOGEE. The histograms inside each plot show the difference between the two samples. The bias ($\mu$), and scatter ($\sigma$) are annotated inside each panel. The bottom row shows histogram distributions for $T_{eff}$ (left), [M/H] (middle), and log \emph{g} (right) of our training sample.} 
	\label{fig:train_apo}
\end{figure*}

Figure \ref{fig:train_apo} shows the comparison between LAMOST training sample (D22) and APOGEE sample (L21). The T$_{eff}$ and [M/H] of training sample have good agreement with APOGEE, which is expected because the parameters are obtained from models trained by APOGEE data. The log \emph{g} distribution has a scatter of 0.16 dex and bias of 0.01 dex. The log \emph{g} is larger than 4.5. The APs distribution of our training sample is shown in the bottom panels of Figure \ref{fig:train_apo}. The distribution of $T_{eff}$ is ranging from 3200 K to 4300 K with a peak at 4000 K. The [M/H] is centered at -0.25 dex, ranging from -1 dex to 0.5 dex. Thus we find that there is no significant difference in the parameters determined by L21 and D22. Since the L21 $T_{eff}$ and [M/H] values are directly tied to the APOGEE scale we adopt these and log \emph{g} values from D22. By selecting signal to noise in i band (SNRi) > 20 and 3$\sigma$ clipping, we obtain a sample of 94,904 LAMOST objects with APs. These APs are the output labels of the models that will be trained.

\section{Model Construction} \label{sec:model}

To estimate the APs of Gaia cool dwarfs, we trained two machine learning models (RF and LightGBM) using photometry in eight bands (r, i, z, J, H, K, W1, and W2) and the APs of LAMOST/APOGEE M dwarfs ($T_{eff}$, log $\emph{g}$, and [M/H]). The feature contribution during model building was analyzed by the SHAP method.

\subsection{Algorithms}

As a part of machine learning, the Random Forest (RF) method integrates multiple decision trees to determine the result of the final vote of each tree \citep{breiman1996bagging, breiman1996out}. It runs efficiently on large databases and can handle thousands of input variables without deleting the variables. With the development of machine learning, RF has extensively worked in different fields as well as in astronomy for APs. For example, \cite{2019AJ....158...93B} predicted the effective temperature for Gaia DR2 data applying RF, with an RMS of 191 K.

In the gradient boosting family of machine learning models, Light Gradient Boosting Machine (LightGBM) was proposed by the Microsoft DMTK team with a performance that exceeds other Boosting Decision Tree tools \citep{ke2017lightgbm}. The histogram algorithm is applied to LightGBM, which occupies less memory and has a lower computational cost. Additionally, LightGBM uses a leaf-wise strategy to grow trees. The leaves with the largest split gain will be found and then split and loop. Compared to levelwise, leafwise can reduce more errors and get better accuracy when the number of splits is the same. However, overfitting will occur when the data size is small. Thus, there are several parameters in the algorithm to avoid overfitting in the training process including learning rate, max$\_$depth (defined depth of the tree), num$\_$leaves (leaf number in tree), lambda$\_$l1 (L1 regularization term) and lambda$\_$l2 (L2 regularization term). A detailed description of LightGBM can be accessed at \url{https://lightgbm.readthedocs.io/en/latest/}. The LightGBM algorithm has been used in the detection of exoplanets \citep{2021MNRAS.tmp.3373M}, searches for cataclysmic variables \citep{2021Univ....7..438H}, forecast of solar flares \citep{2021AC....3500468R} as well as the prediction of stellar atmospheric parameters \citep{2022AJ....163..153L}.

\subsection{Feature Testing}

In order to decide whether data features should use magnitudes or colors, the sample was divided into training and testing data with fractions of 0.8 and 0.2, respectively. Table \ref{tab:methods} lists the bias and scatter between the predicted and true APs of testing data set for both RF and LightGBM (LGB). For the $T_{eff}$ model, the $\sigma$ of RF and LGB using colors is less than 70 K which is better than that using magnitudes (see Table 1). The colors are more sensitive to $T_{eff}$, which is also actually known and has been applied to estimate $T_{eff}$ of different objects \citep{2021MNRAS.507.2684C}. Nonetheless, for the [M/H] and log \emph{g}, the use of colors or magnitudes provides approximate the same scatter and bias. In order to be self-consistent with the $T_{eff}$, colors are also applied to be the features for the estimation of [M/H] and log \emph{g} values. By comparing the performance of test data, colors (r - i, i - z, z - J, J - H, H - K, K - W1, W1 - W2) were adopted for use as the data features. 

\begin{table}[h]
\centering
\caption{Testing results of the two methods using both features of magnitudes and colors. Here, the magnitudes are absolute for extinction in r, i, z, J, H, K, W1, and W2 bands, and the colors are r - i, i - z, z - J, J - H, H - K, K - W1, and W1 - W2. The $\mu$ and $\sigma$ present the bias and scatter of the sample.}
\label{tab:methods}
\setlength{\tabcolsep}{3mm}{
\begin{tabular}{cccccc}
    \hline
    \hline
    &\multicolumn{4}{c}{feature = magnitude}\\
    \hline
    &\multicolumn{2}{c}{RF}  &\multicolumn{2}{c}{LGB}   \\
    &$\mu$& $\sigma$ & $\mu$ & $\sigma$\\
    \hline
    $T_{eff}$ & 0.441  & 72 & -5.14  & 80&\\
    $[M/H]$ & -0.004  & 0.22 & -0.006 & 0.23& \\
    log $\emph{g}$  & -0.003 & 0.054 & -0.003 & 0.057&\\
    \hline
    \hline
    & \multicolumn{4}{c}{feature = color}\\
    \hline
    & \multicolumn{2}{c}{RF}  & \multicolumn{2}{c}{LGB} \\
    & $\mu$ & $\sigma$  & $\mu$  & $\sigma$\\
    \hline
    $T_{eff}$  & -0.5 & 69  & 5.7 & 68&\\
    $[M/H]$  & 0.004 & 0.23  & 0.006 & 0.22&\\
    log $\emph{g}$ & 0.0 & 0.054 & 0.003 & 0.053&\\
    \hline
\end{tabular}}
\end{table}

\begin{figure*}[htbp]
	\centering
	\includegraphics[width=0.95\textwidth]{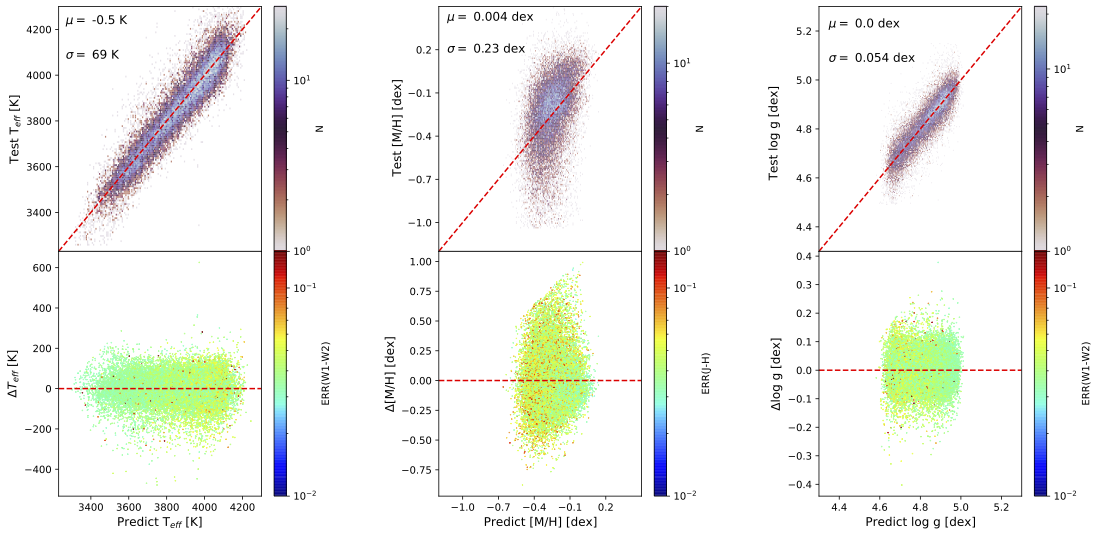}
	\caption{Comparison diagrams of the testing samples for true and predicted value in terms of $T_{eff}$ (left), [M/H] (bottom), and log \emph{g} (right). The top panels show the comparisons of predicted values from RF and true values, color-coded by the number density of the test samples. The bottom panels indicate the difference between predicted and true value, color-coded by the photometric error of color W1 - W2 for $T_{eff}$ and log \emph{g}, color J - H for [M/H].}
	\label{fig:test}
\end{figure*}

Using colors as features, Figure \ref{fig:test} shows the comparison of predicted and true values of the testing data. The APs estimated by RF show a similar result to those with LGB. In the top panels, the predicted $T_{eff}$ has a good agreement with that of the testing sample, having a bias of -0.5 K scatter of 69 K. The density distribution of [M/H] has a slightly larger scatter of 0.2 dex. The log \emph{g} distribution has a scatter of 0.05 dex without systematic bias. The residuals of predicted and true values is shown in the bottom panels of Figure \ref{fig:test}, with the photometric color error of W1 - W2 for $T_{eff}$ and log \emph{g}, color J - H for [M/H]. The points with different photometric errors are randomly distributed around the residual APs. Thus, the effect of the photometric error on estimating APs is not considered further.

\subsection{Feature Importance Analysis}

It is instructive to analyze the importance of features on the APs of cool dwarfs. We applied a Python package $\emph{shap}$ to compute the SHAP (Shapley Additive ExPlanations) value of each feature \citep{NIPS2017_7062}, which can create an explanation for feature contribution to the model. The SHAP values of each feature are calculated and plotted in a SHAP summary graph. The vertical axis shows features ordered by their importance for modeling. The color is the feature value of the training sample. The horizontal axis indicates the SHAP value of each feature, which is positive and negative. The larger the absolute SHAP value of a feature, the larger the influence of this feature during model building.

\begin{figure}[h]
	\centering
	\includegraphics[scale=0.3]{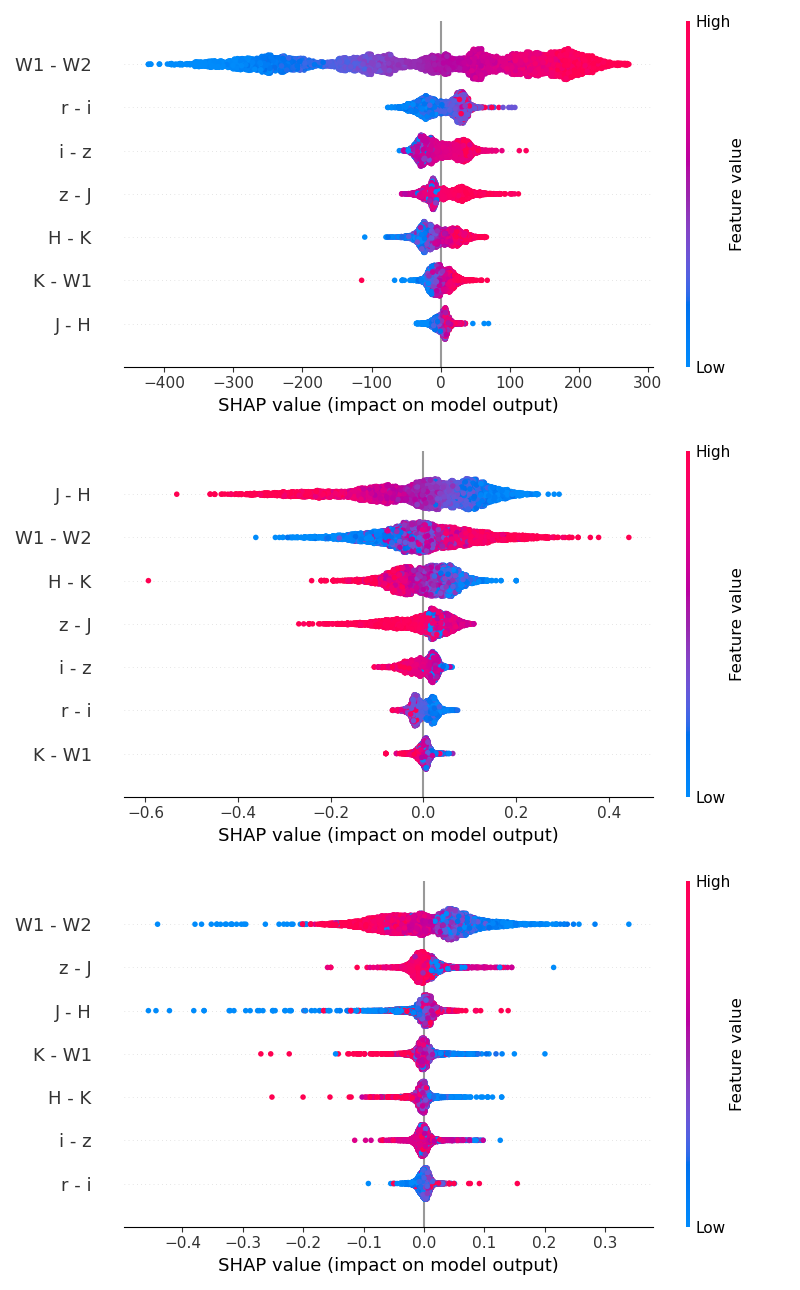}
	\caption{Feature importance graph during model building for $T_{eff}$ (top), [M/H] (middle) and log $\emph{g}$ (bottom). The color is the feature value, the x-axis is the SHAP value and the y-axis for each feature is ordered by their contribution to model construction.}
	\label{fig:shap}
\end{figure}

We find similar results from both the RF and LightGBM methods, here we only present the SHAP graph for RF in terms of $T_{eff}$, [M/H], and log $\emph{g}$ in Figure \ref{fig:shap}. The top panel is the SHAP graph for $T_{eff}$ model, color-coded by the feature values. The color W1 - W2 has the largest positive relationship to the SHAP value, which means W1 - W2 is the most sensitive to variations in $T_{eff}$. In the middle panel of the [M/H] model, the near-infrared bands play a more important role during model building. 

The three most important colors J - H, W1 - W2, and H - K are all in the infrared bands, indicating that the infrared photometry is key to estimate the [M/H]. \cite{2016MNRAS.460.2611S} has also explored that the colors J - K and W1 - W2 are promising metallicity indicators for the $T_{eff}$ and log $\emph{g}$ of the M dwarfs. In the log $\emph{g}$ SHAP graph, the color W1 - W2 is the most effective feature, followed by the z - J and J - H. The three most important features for APs estimation are summarized in Table \ref{tab:shap}. It can be found that the near- and mid-infrared photometry is more sensitive to the stellar APs variation. The importance of infrared photometry for M dwarfs has long been known, e.g., \cite{10.1093/mnras/171.1.19P} introduced that M dwarfs in JHK colors are quite distinct from M giants cause the dominated water opacity. \cite{2022A&A...657A.129A} found that the H broad band is the most relevant feature for estimation of log \emph{g} of M dwarfs using near-infrared spectroscopy. Thus our finding of the significance of infrared photometry for the determination of APs for M dwarfs is in line with expectations.

\begin{table}[h]
    \centering
    \caption{The three most important features that are sensitive to $T_{eff}$, [M/H], and log \emph{g} during modeling process.}
    \setlength{\tabcolsep}{5mm}{
    \begin{tabular}{lccc}
    \hline
    \hline
    & $T_{eff}$ & [M/H] & log \emph{g} \\
    \hline
    First & W1 - W2 & J - H & W1 - W2 \\
    Second & r - i & W1 - W2 & z - J \\
    Third & i - z & H - K & J - H \\
    \hline
    \end{tabular}}
    
    \label{tab:shap}
\end{table}

\section{Parameterization for Gaia Cool Dwarfs}
\label{sec:catalog}

In this section, we estimated the APs of Gaia DR3 cool dwarfs using the trained RF and LightGBM models in Section \ref{sec:model}. We provided a catalog of APs for 1,806,921 cool dwarfs in Gaia DR3. We compared our estimated APs to those provided by the Gaia DR3 release based on Gaia BP/RP spectra and RVS spectra.

\subsection{AP Catalog from Photometry}

In this section, we determin the APs for Gaia cool dwarfs using the models trained in Section \ref{sec:model}. We provided a stellar AP catalog of 1,806,921 cool dwarfs, containing multiband photometry, $T_{eff}$, [M/H], log $\emph{g}$ by RF and LightGBM and the corresponding uncertainties. Here, the uncertainty is derived from the multiple recalculations based on the Monte Carlo (MC) method. We trained 100 different models using randomly selected sub-training samples. Then the 100 APs for each object are predicted by repeated ML models applied to compute the dispersion as the uncertainty of this object. The field descriptors of the catalog are listed in Table \ref{tab:field} and an example section of the catalog shown in Table \ref{tab:catalog}. The complete catalog is accessible at \url{https://nadc.china-vo.org/res/paperdata/data_upload?resource=101203}.

\begin{table}[htbp]
\scriptsize
% 	\centering
	\caption{Observed and derived parameters for our catalog of Gaia cool dwarfs.}
	\label{tab:field}
	\begin{threeparttable}
	\setlength{\tabcolsep}{0mm}{
	\begin{tabular}{llc}
	     
% 	\begin{tabular}{8cm}{llp{4cm}}
		\hline
		\hline
		Column & Unit & Description \\
		\hline
		Gaia DR3 SourceID & & Gaia Identification ID\\
		ra$\_$obs & deg & right ascension of object\\
		dec$\_$obs & deg & declination of object\\
		rmag & mag & SDSS r band magnitude\\
		imag & mag & SDSS i band magnitude\\
		zmag & mag & SDSS z band magnitude\\
		Jmag & mag & 2MASS J band magnitude\\
		Hmag & mag & 2MASS H band magnitude\\
		Kmag & mag & 2MASS K band magnitude\\
		W1mag & mag & ALLWISE W1 band magnitude\\
		W2mag & mag & ALLWISE W2 band magnitude\\
		TEFF$\_$RF & K & Effective temperature from RF\\
		TEFF$\_$RF$\_$ERR & K & Uncertainty of Teff from RF\\
		$\left [ M/H \right ]\_$RF & dex & Metal abundance from RF\\
		$\left [ M/H \right ]\_$RF$\_$ERR & dex & Uncertainty of [M/H] from RF\\
		LOGG$\_$RF & dex & Surface gravity from RF\\
		LOGG$\_$RF$\_$ERR & dex & Uncertainty of log $\emph{g}$ from RF\\
		TEFF$\_$LGB & K & Effective temperature from LightGBM\\
		TEFF$\_$LGB$\_$ERR & K & Uncertainty of Teff from LightGBM\\
		$\left [ M/H \right ]\_$LGB & dex & Metal abundance from LightGBM\\
		$\left [ M/H \right ]\_$LGB$\_$ERR & dex & Uncertainty of [M/H] from LightGBM\\
		LOGG$\_$LGB & dex & Surface gravity from LightGBM\\
		LOGG$\_$LGB$\_$ERR & dex & Uncertainty of log $\emph{g}$ from LightGBM\\
            flag & & boundary of BP - RP = 2.9\\
		\hline
	\end{tabular}}
	
	\begin{tablenotes}
		\footnotesize	
		\item[1] The photometry in each band is the absolute magnitude for reddening and extinction.
		\item[2] The uncertainties of APs are the dispersion of 100 predicted values, from models trained by 100 randomly selected sub$\_$training samples.
            \item [3] The flag is the boundary of BP - RP = 2.9 mag, flag = 1 while BP - RP > 2.9, otherwise flag = 0. 
	\end{tablenotes}
	\end{threeparttable}
\end{table}

\begin{table*}
	\centering
	\caption{The stellar atmospheric parameter catalog of Gaia cool dwarfs.}
	\label{tab:catalog}
	\resizebox{0.9\textwidth}{!}{
	\begin{threeparttable}{
	\begin{tabular}{cccccccc}
		\hline
		\hline
		Gaia DR3 SourceID & TEFF$\_$RF & $\left [ M/H \right ]\_$RF & LOGG$\_$RF & TEFF$\_$LGB  & $\left [ M/H \right ]\_$LGB & LOGG$\_$LGB\\ 
		\hline
        352221678055936 & 3950 & -0.12 & 4.80 & 3936 & -0.23 & 4.80 \\
        354523780537216 & 3662 & -0.34 & 4.92 & 3645 & -0.40 & 4.91 \\
        357792251311232 & 3653 & -0.14 & 4.93 & 3618 & -0.18 & 4.93 \\
        360541030376576 & 4180 & -0.29 & 4.67 & 4157 & -0.26 & 4.64 \\
        361296944678016 & 3837 & -0.01 & 4.81 & 3827 & -0.02 & 4.84 \\
        363014931564416 & 4180 & -0.36 & 4.64 & 4073 & -0.35 & 4.66 \\
        363461608162048 & 3667 & -0.4 & 4.91 & 3656 & -0.38 & 4.91 \\
        363495967900032 & 3771 & -0.32 & 4.87 & 3770 & -0.32 & 4.86 \\
        363560391977856 & 3417 & -0.12 & 4.99 & 3438 & -0.13 & 4.96 \\
        367138100139136 & 3813 & -0.29 & 4.86 & 3813 & -0.23 & 4.84& \\
        \vdots\\
        \hline
	\end{tabular}}
	\footnotesize
	{The complete table can be accessed at $https://nadc.china-vo.org/res/paperdata/data_upload?resource=101203$}
	\end{threeparttable}}

\end{table*}

\begin{figure*}[htbp]
    \centering
    \includegraphics[scale=0.5]{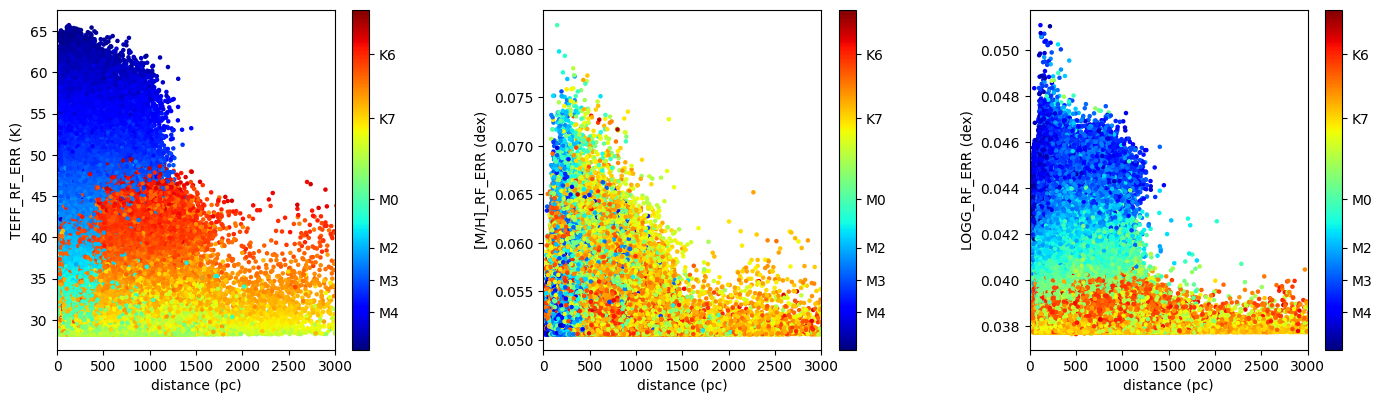}
    \caption{Density distribution of uncertainty of $T_{eff}$ (left), [M/H] (middle), and log \emph{g} (right) along with the distance, color-coded by spectra type. Here, the error of APs is the dispersion of 100 predicted values by repeating modeling using 100 sub-training samples.}
    \label{fig:error}
\end{figure*}

In the catalog, the uncertainties of our APs are provided to demonstrate the computed error. The density distribution of uncertainty for $T_{eff}$, [M/H], and log \emph{g} as a function of the distance is shown in Figure \ref{fig:error}. These stars are within 3 kpc, and an uncertainty of $T_{eff}$, [M/H], log $\emph{g}$ less than 65 K, and 0.08 dex, and 0.05 dex, respectively. Objects with larger uncertainties are the later M spectra types, which is not surprising since this is where the training set is comparatively smaller. 

\begin{figure*}[htbp]
	\centering
	\includegraphics[width=0.9\textwidth]{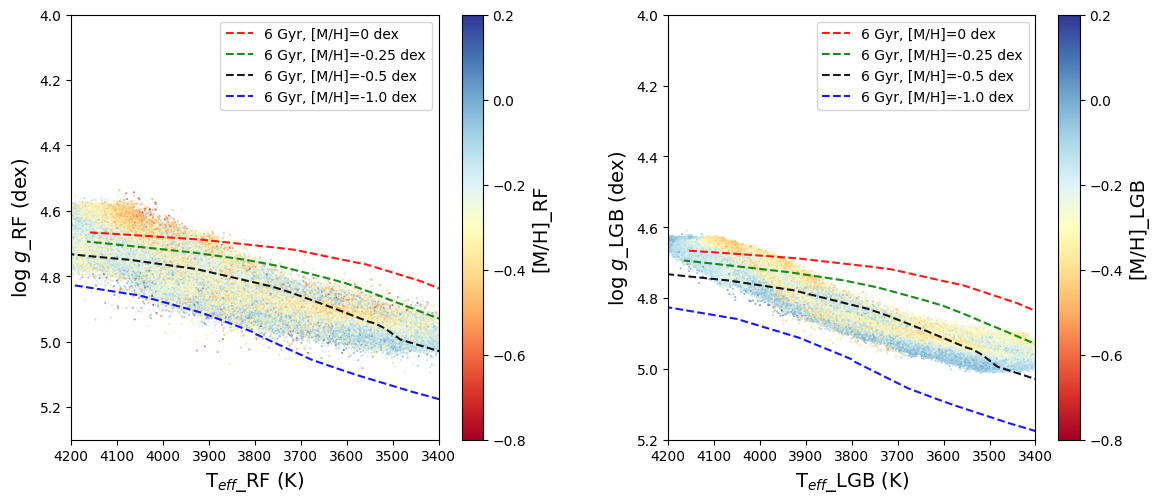}
	\caption{Kiel diagram of our catalog: log \emph{g} as a function of $T_{eff}$ color-coded by [M/H]. The solid lines indicate PARSEC isochrones with an age of 6 Gyr and various [M/H]s. The left panel shows the HRD of our catalog predicted by RF, plotting four isochrones with the age of 6 Gyr and [M/H] of 0, -0.25, -0.5, and -1.0 dex. The right panel indicates the HRD of our catalog predicted by LGB, with the isochrones the same as the left panel.}
	\label{fig:parsec}
\end{figure*}

Figure \ref{fig:parsec} presents the Kiel Diagram of our catalog. We also showed the PARSEC (PAdova and TRieste Stellar Evolution Code) theoretical tracks, at an age of 6 Gyr and different [M/H] values from PARSEC version 1.2S \citep{2012MNRAS.427..127B, 2015MNRAS.452.1068C}. The points in the left panel (predicted by RF) are overlap with four isochrones with [M/H] = 0, -0.25, -0.5, and -1.0 dex, and the same isochrones are in the right panel (predicted by LGB). It can be seen that the APs of LGB are less scattered than those of the RF. Nonetheless, the greater scatter of the LGB APs may actually be more physically realistic given the expected scatter in the APs of the objects.

\subsection{Parameter Comparisons}

In this section, we compared our estimated APs with three catalogs, including APs from high- and low- resolution spectra of Gaia DR3, and APOGEE spectra.

\subsubsection{Comparison with APs from APOGEE spectra}

The Apache Point Observatory Galactic Evolution Experiment (APOGEE) is one of the parts of SDSS. The stellar APs and metal abundances are determined by the APOGEE Stellar Parameters and Abundances Pipeline (ASPCAP), which analyzes the spectra of the APOGEE targets \citep{2018AJ....156..125H, 2006AJ....131.2332G}. The ASPCAP first obtains stellar parameters by fitting the entire spectrum, and second, individual element abundance is decided by fitting a limited spectrum associated with the element. ASPCAP is known to have the potential to allow eight parameters, including effective temperature $T_{eff}$, surface gravity log $\emph{g}$, etc. We cross-matched our sample with APOGEE DR16 stellar AP catalog, and selected cool dwarfs by following these criteria:

\begin{enumerate}
    \item \emph{ASPCAPFLAG = 0} and \emph{STARFLAG = 0}.
    \item S/N of APOGEE > 30.
    \item $T_{eff}\_$error < 150 K, log $\emph{g}\_$error < 0.08 dex, and [M/H]$\_$error < 0.02 dex.
    \item 3$\sigma$ clipping for $T_{eff}$, [M/H], and log $\emph{g}$.
    \item 3200 K < $T_{eff}$ < 4200 K and log $\emph{g}$ > 4.5 dex.
\end{enumerate}

\begin{figure}[htbp]
    \centering
    \includegraphics[width=0.4\textwidth]{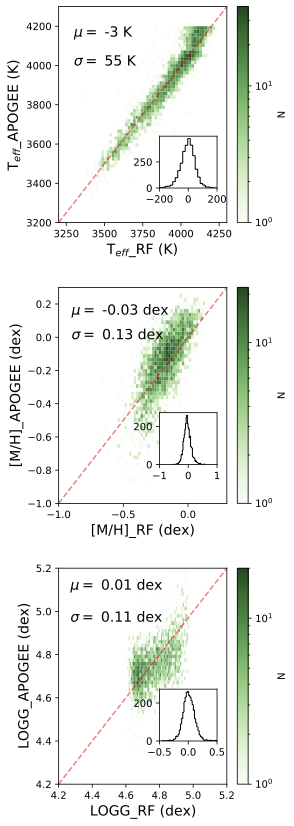}
    \caption{APs comparison between APOGEE DR16 and the RF values for this work for $T_{eff}$ (top), [M/H] (middle), and log \emph{g} (bottom). The difference in histogram between this work and APOGEE is plotted inside each panel. The x-axis represents the residual, and the y-axis represents the number of occurrences.}
    \label{fig:esto_apo}
\end{figure}

Figure \ref{fig:esto_apo} shows the predicted APs compared with APOGEE DR16. As both RF and LGB yield similar results, we only displayed the APs from RF for comparison with APOGEE DR16 data. The $T_{eff}$ distribution has a good agreement, with a bias of -3 K and a scatter of 55 K. This agreement was expected since our training sample was labeled by APOGEE high-resolution spectra. The [M/H] distribution has a bias of -0.03 dex and a scatter of 0.13 dex, while the log \emph{g} estimation has a bias of 0.01 dex and a scatter of 0.11 dex. Overall, the APs of this work are in good agreement with APOGEE DR16.

\subsubsection{Comparison with APs from BP/RP spectra}

\begin{figure}
    \centering
    \includegraphics[width=0.45\textwidth]{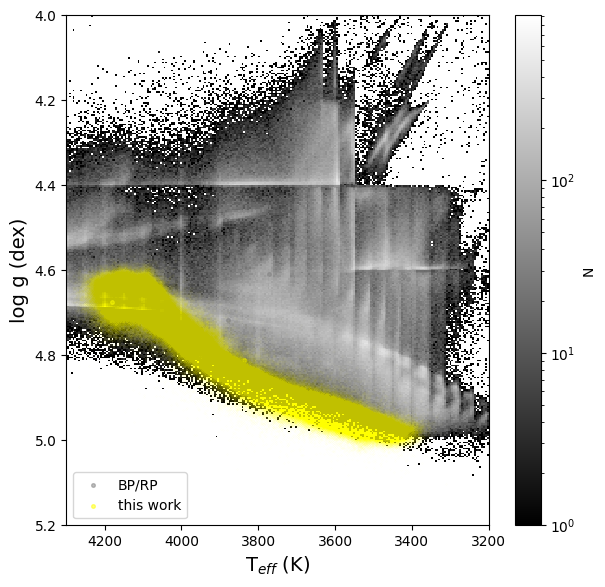}
    \caption{APs comparison between RF values from our work (yellow points) and DR3 APs from BP/RP spectra (grey points) in terms of log $\emph{g}$ and $T_{eff}$.}
    \label{fig:esti_photo}
\end{figure}

We performed a cross-match of Gaia DR3 APs from BP/RP spectra, APs with our catalog in TOPCAT and excluded sources without APs. Additionally, we selected objects of $T_{eff}$ < 4300 K, log \emph{g} > 4 dex, and [M/H] from -1.0 to 0.1 dex to ensure a valid comparison. A total of 1,599,278 objects were included and displayed in Figure \ref{fig:esti_photo}. The yellow points represent RF APs from our work BP/RP spectra, while the grey points represent the results from Gaia DR3. The grey points suggest a wide variety of artefacts in the Gaia DR3 APs which appear to be unphysical. In contrast, our work provides estimations for cool dwarfs which appear to be much more in line with those expected based on models, e.g., in Figure \ref{fig:parsec}.

\subsubsection{Comparison with APs from RVS spectra}

\begin{figure}[h]
	\centering
	\includegraphics[width=0.45\textwidth]{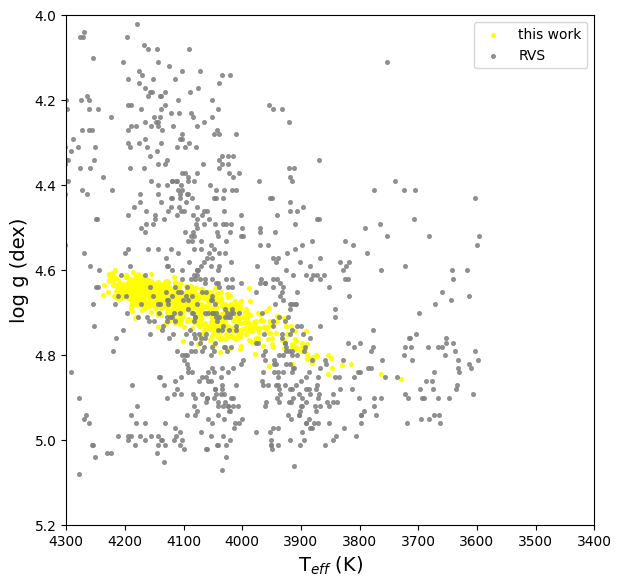}
	\caption{APs comparison of this work with RVS spectra in terms of log $\emph{g}$ and $T_{eff}$. The yellow points represent this work, while the grey points depict RVS spectra.}
	\label{fig:esti_spec}
\end{figure}

In addition to the APs from low-resolution BP/RP spectra, Gaia DR3 also provides the APs from high-resolution spectra. In this study, we select common objects between RVS and our catalog. After performing the matching process, we were left with 839 objects for comparison. Figure \ref{fig:esti_spec} illustrates the APs comparison between RVS and this work. It is immediately obvious from the figure that there are significant discrepancies relative to our expectations based for example on the models in Figure \ref{fig:parsec}. The differences here can be attributed to the accuracy of log \emph{g} decreasing as $T_{eff}$ decreases \citep{2022arXiv220605541R}.

The $\sigma$ could be increasing in generating APs, which is probably caused by two reasons. On the one hand, the machine learning method prefers to trend the high distribution of the training sample. The bigger scatter would be exist if the parameter distribution differ to the training sample. On the other hand, the APs of cools dwarfs from Gaia need to be careful while using them. The limitations cannot be avoided, such as the low resolution of spectra, deficiency of wavelength band, and the lack of template of cool dwarfs. 

\section{Summary}\label{sec:summary}

In this work, we developed estimation models for $T_{eff}$, [M/H], and log \emph{g} for cool dwarfs by applying machine learning methods to optical and infrared photometric data. The main contributions of this paper are summarized as follows.

\begin{itemize}
    \item
    We selected 94,904 objects from the stellar atmospheric parameter catalogs of LAMOST and APOGEE M dwarfs, with $T_{eff}$ ranging from 3200 K to 4300 K, [M/H] ranging from -0.1 dex to 0.5 dex, and log $\emph{g}$ larger than 4.5 dex. Using multi-band photometry of optical and infrared surveys, two machine learning methods (RF and LightGBM) were applied to construct models to predict stellar atmospheric parameters. The parameters showed relative little sensitivity to the chosen machine learning method and present a best dispersion of 68 K, 0.22 dex, and 0.053 dex for $T_{eff}$, [M/H], and log $\emph{g}$, respectively.
    \item
    We used the SHAP values to analyze the importance of the features during model building for the APs. According to the SHAP value diagram, it is found that color W1 - W2 is the most sensitive feature for both $T_{eff}$ and log \emph{g}, and that J - H color is most important for [M/H]. Infrared photometry plays a crucial role in building our APs models.
    \item   
    On the basis of the above work, we have presented a stellar atmospheric parameter catalog of Gaia DR3 cool dwarfs. There are a total of 1,806,921 objects with photometry in eight bands and stellar atmospheric parameters ($T_{eff}$, [M/H], and log $\emph{g}$), as well as their corresponding uncertainties. Comparing our catalog APs with those from Gaia DR3 BP/RP spectra and RVS suggest a number of issues with the Gaia DR3 APs for cool dwarfs.
\end{itemize}

\section*{ACKNOWLEDGEMENTS}
This work is supported by the National Key R\&D Program of China No. 2019YFA0405502, the National Natural Science Foundation of China (12103068, U1931209), a Visiting Fellowship from the Alliance of International Science Organizations, and science research grants from the China Manned Space Project with NO.CMS-CSST-2021-B05. 

This research has made use of LAMOST data \citep{2015RAA....15.1095L}. The full name of LAMOST is the Large Sky Area Multi-Object Fiber Spectroscopic Telescope or Guoshoujing Telescope. It is a National Major Scientific Project built by the Chinese Academy of Sciences. Funding for the project has been provided by the National Development and Reform Commission. LAMOST is operated and managed by the National Astronomical Observatories, Chinese Academy of Sciences.

This research makes use of data products from \textit{Wide-field Infrared Survey Explorer} \citep{2010AJ....140.1868W}, which is a joint project of the University of California, Los Angeles, and the Jet Propulsion Laboratory/California Institute of Technology, funded by the National Aeronautics and Space Administration. 

This work has made use of data from the European Space Agency (ESA) mission Gaia (https://www.cosmos.esa.int/gaia), processed by the Gaia Data Processing and Analysis Consortium (DPAC, https://www.cosmos.esa.int/web/gaia/dpac/consortium). Funding for the DPAC has been provided by national institutions, in particular, the institutions participating in the Gaia Multilateral Agreement.

The data products from the Two Micron All Sky Survey \citep{2006AJ....131.1163S}, which is a joint project of the University of Massachusetts and the Infrared Processing and Analysis Center/California Institute of Technology, funded by the National Aeronautics and Space Administration and the National Science Foundation.

This work makes use of data from SDSS DR16 of the SDSS-IV \citep{2017AJ....154...28B}. Funding for SDSS-IV has been provided by the Alfred P. Sloan Foundation, the Participating Institutions, the National Science Foundation, and the US Department of Energy Office of Science.

This research uses Astropy\footnote{\url{http://www.astropy.org}}, TOPCAT \citep{2005ASPC..347...29T}, Scipy \citep{2020NatMe..17..261V}, Numpy \citep{2020Natur.585..357H}.

%%%%%%%%%%%%%%%%%%%%%%%%%%%%%%%%%%%%%%%%%%%%%%%%%%

\bibliography{bibfile}{}
\bibliographystyle{aasjournal}

% \begin{appendix}
% \setcounter{figure}{}
% \section{figure}
% \setcounter{figure}{0} 

% \end{appendix}
%%%%%%%%%%%%%%%%%%%% REFERENCES %%%%%%%%%%%%%%%%%%

% The best way to enter references is to use BibTeX:

% Don't change these lines

% \label{lastpage}

%% This command is needed to show the entire author+affiliation list when
%% the collaboration and author truncation commands are used.  It has to
%% go at the end of the manuscript.
%\allauthors

%% Include this line if you are using the \added, \replaced, \deleted
%% commands to see a summary list of all changes at the end of the article.
%\listofchanges
\end{document}